\documentclass[12 pt]{article}
\usepackage{amsmath, amsfonts, amssymb}
\usepackage{float}
\usepackage{graphicx}
\usepackage{indentfirst}
\usepackage{threeparttable}
\usepackage{url}
\usepackage[english]{babel}
\usepackage{xcolor}
\usepackage[left=2cm,right=2cm,
    top=2cm,bottom=2cm,bindingoffset=0cm]{geometry}

\begin{document}

\title{\LARGE \bf Hyperfine transitions in atoms and rotational transitions in molecules of the early Universe \\as a possible source of CMB distortions}
\author{\bf M.N.~Golubev$^1$\thanks{E-mail: golubew.maxim2015@yandex.ru}  
,
P.A.~Kislitsyn$^2$, A.V. Ivanchik$^2$}
\date{\it  \small  $^1$ Peter the Great St.\,Petersburg Polytechnic University, St.\,Petersburg, Russia \\
\it  \small  $^2$ Ioffe Institute, St.\,Petersburg, Russia }

\maketitle

\renewcommand{\abstractname}{}
\begin {abstract}
\bf 
In the near future, measurements are expected to detect the influence of the hyperfine transition in the ground state of the hydrogen atom on the cosmic microwave background (CMB) spectrum. This raises the question of what other substances could produce similar distortions and to what extent.
This work presents a systematic analysis of the potential for the most abundant atoms, molecules, and their ions to distort the CMB spectrum. In addition, the wavelength ranges for all the transitions considered in which such distortions might be observed have been calculated. As a result, we conclude that the distortions caused by the most abundant elements are orders of magnitude smaller than those produced by hydrogen.
\rm

\end{abstract}


\section{Introduction}

During the Dark Ages, neutral hydrogen absorbed and emitted photons at $1420\, \mathrm{MHz}$.The rate of absorption and emission depended on the temperature difference between the hydrogen and the CMB. The approximate spectra (Pritchard, 2012) of this distortion are shown in Figure 1.

\begin{figure}[H]
\includegraphics[width=\textwidth]{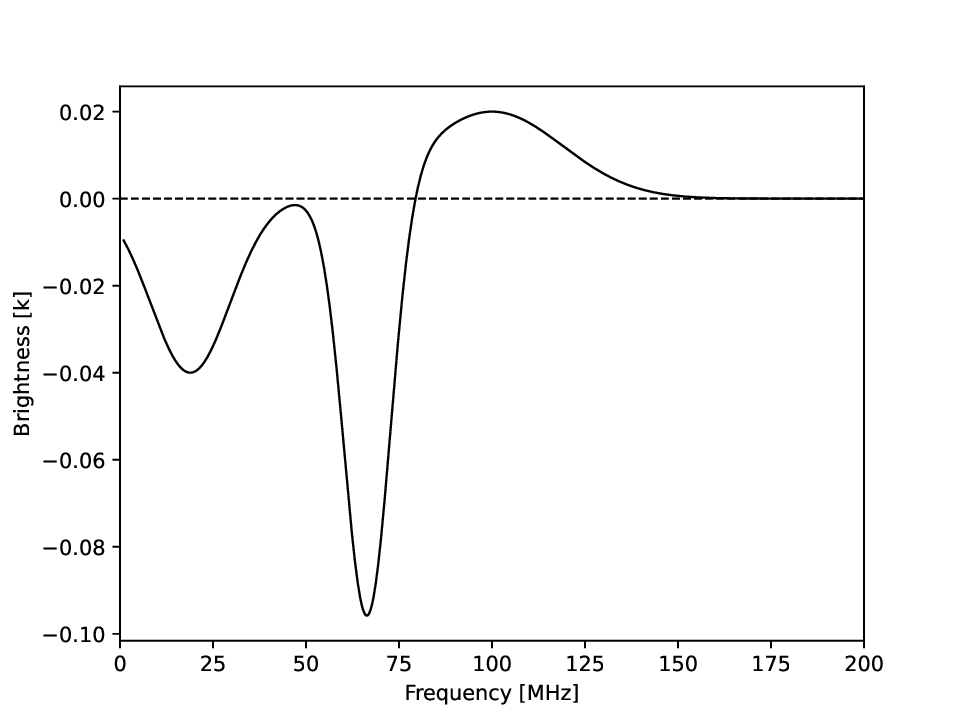}
\begin{center}
\caption{ The schematic representation of spectral distortion brightness temperature. The dashed line corresponds to the zero-reference level
$T_b = 0$ ($T_{observed} = T_{CMB}$) }
\end{center}
\end{figure}

Since no known sources of electromagnetic radiation existed during this epoch, these distortions offer crucial insights into the Dark Ages—particularly into the thermal state of baryonic matter. The baryonic temperature provides valuable constraints on potential heating mechanisms, such as primordial black holes, cosmic strings, and other exotic phenomena, while also helping to infer star formation rates at high redshifts. However, these distortions remain undetected to date due to their exceptionally weak signal amplitude (Pritchard, 2012).

\section{Relative amplitude calculation of spectral distortions}

The influence of the 1420 MHz line can be used to study the properties of structures of neutral hydrogen, essentially representing the state during the Dark Ages. To model this interaction, consider the radiative transfer equation:

\begin{equation}
        \frac{dI_{\nu}}{ds} = -\alpha_{\nu}I_{\nu}+j_{\nu}
\end{equation}

Where $I_{\nu}$ is the intensity at frequency $\nu$, $s$ is a coordinate along a line of sight and $\alpha_{\nu}$, $j_{\nu}$ are the absorption and emission coefficients. Within the Rayleigh-Jeans approximation, the intensity of radiation can be expressed in terms of the brightness temperature of the cosmic microwave background (CMB). Under this assumption, the solution to the radiative transfer equation (1) takes the form:

\begin{equation}
        T_{obs}(\nu) = T_{ex}(1-e^{-\tau_{\nu}}) +T_{R}e^{-\tau_{\nu}}
\end{equation}

In this equation $ T_{obs}$ is the observed brightness temperature, $T_{ex}$ is the CMB temperature, $\tau$ is optical opacity, $T_{R}$ is brightness temperature of the source. In this case, it is the spin temperature of neutral hydrogen - $T_s$. The opacity is given by the following expression (Pritchard,2012):

\begin{equation}
        \tau_{\nu} = \int ds\left(1-e^{-\frac{E}{kT_s}}\right)\sigma_o \phi(\nu)n_0
\end{equation}

Where $n_0$ is the number density of atoms in the lower energy level, $E$ is the energy of the transition, $\sigma_o$ is the transition cross-section at line center, $\phi(\nu)$ is spectral line profile and $k$ is the Boltzman constant. Since, in addition to collisions between gas particles, interactions with CMB and $Ly\alpha$ photons also occur, the distribution across hyperfine energy levels is governed not only by the kinetic, but by the spin temperature, which is given by the expression(Pritchard,2012):

\begin{equation}
        T_s^{-1} = \frac{T_{CMB}^{-1}+x_\alpha T_\alpha^{-1} + x_kT_k^{-1}}{1+x_\alpha +x_k}
\end{equation}

Where $T_\alpha$ is the color temperature of $Ly\alpha$ radiation field, $T_k$ is the kinetic temperature of gas, $x_k$ and $x_\alpha$ are the coupling coefficients. Taking into account that the opacity is small, we can expand (2) into a Taylor series and get the following:

\begin{equation}
        \delta T_b = \frac{T_s - T_{CMB}}{1+z}\tau
\end{equation}

Where z is the cosmological redshift. To evaluate the relative amplitude of distortions, we will assume that the spin temperatures of all elements are identical and that the spectral line profiles are a delta function: 

\begin{equation}
        \frac{\delta T_{bi}}{\delta T_{bj}} = \frac{\frac{T_{si} - T_{CMB}}{1+z}\tau_i}{\frac{T_{sj} - T_{CMB}}{1+z}\tau_j} = \frac{\tau_i}{\tau_j}=\frac{(1-exp(-\frac{E_i}{kT_s}))\frac{A_i}{\nu_i^2}n_{0i}}{(1-exp(-\frac{E_j}{kT_s}))\frac{A_j}{\nu_j^2}n_{0j}} = \alpha
\end{equation}
This quantity ratio will be further designated as $\alpha$.

\section{Relative amplitude of distortion}
\subsection{Atoms}
Brightness temperature fluctuation was non-zero from z ~200 to z ~6.5 (Pritchard,2012).  During this epoch, the cosmic gas temperature was sufficiently low to prevent excitation to higher electronic energy states, ensuring atoms remained in their ground states. Hyperfine transitions require atoms and ions with non-zero nuclear spin. Thus, helium-4 and its ion will not make distortion because theirs nuclear spin equal zero. Additionally, if all hyperfine levels are occupied by electrons, transition is impossible due to the Pauli exclusion principle.This rules out contributions from neutral helium-3 and lithium ions. The relative amplitudes of distortion for the most abundant atoms and their ions are presented in Table 1.

\begin{table}[h]
\centering
\caption{Hyperfine transitions in atoms: Wavelengths, Einstein coefficients, relative abundances at \( z = 10 \), and dimensionless parameter \( \alpha \) (defined in eq. (6)). References are provided for each quantity.}
\label{tab:hyperfine_transitions}
\begin{tabular}{|l|c|c|c|c|}
\hline
\textbf{Atom/Ion} & \textbf{\( \lambda \) (cm)} & \textbf{\( A_{ul} \) (s\(^{-1}\))} & \textbf{\( \frac{n}{n_H} \)} & \textbf{\( \alpha \)} \\ 
\hline 
H                 & 21.1  & \(2.87 \times 10^{-15}\) & 1                  & 1                  \\
D                 & 91.6  & \(4.69 \times 10^{-17}\) & \(1.0 \times 10^{-5}\) & \(1.0 \times 10^{-6}\) \\
\(^3\)He\(^+\)    & 3.45  & \(1.95 \times 10^{-12}\) & \(1.0 \times 10^{-27}\) & \(1.0 \times 10^{-25}\) \\
\(^6\)Li          & 132   & \(1.59 \times 10^{-17}\) & \(5.0 \times 10^{-13}\) & \(2.0 \times 10^{-14}\) \\
\(^7\)Li          & 37.3  & \(7.79 \times 10^{-16}\) & \(8.0 \times 10^{-12}\) & \(6.0 \times 10^{-12}\) \\
\hline
\end{tabular}
\end{table}

\vspace{5mm}
\footnotesize
\textbf{References:} \\
1. Einstein coefficients and transition widths: Wiese (2009). \\
2. Relative abundances at \( z = 10 \): Galli (2013). \\

\subsection{Molecules}
During the epoch under consideration, the thermal energy in the Universe was sufficient to populate rotational energy levels of molecules but insufficient to excite vibrational energy levels. In Table 2, we list the transition with the largest relative amplitude (compared to hydrogen) for each molecule and its ion, assuming $T_s =300 K$. For molecules lacking published Einstein coefficients, we calculated these values using the standard formula for electric dipole transitions (Hilborn, 1982):
\begin{equation}
       A_{ik}=\frac{64\pi^4}{3hc^3}\nu^3|\vec{P_{ik}}|^2
\end{equation}
Where $A_{ik}$ is the Einstein coefficient of transition, $\nu$ is frequency of transition and $\vec{P_{ik}}$ is dipole moment of molecule.

\begin{table}[h]
\centering
\caption{Rotational transitions in molecules: Wavelengths, Einstein coefficients, relative abundances at \( z = 10 \), and dimensionless parameter \( \alpha \) (defined in eq. (6)). References are provided for each quantity.}
\label{tab:molecular_transitions}
\begin{tabular}{|l|c|c|c|c|}
\hline
\textbf{Molecule} & \textbf{\( \lambda \) (cm)} & \textbf{\( A_{ul} \) (s\(^{-1}\))} & \textbf{\( \frac{n}{n_H} \)} & \textbf{\( \alpha \)} \\ 
\hline 
H\(_2\)          & \(4.4 \times 10^{-4}\) & \(2.9 \times 10^{-11}\) & \(6.3 \times 10^{-7}\)  & \(1.8 \times 10^{-10}\)  \\
H\(_2^+\)        & \(5.6 \times 10^{-4}\) & \(1.6 \times 10^{-10}\) & \(9.2 \times 10^{-15}\) & \(1.6 \times 10^{-15}\) \\
HD               & \(5.9 \times 10^{-4}\) & \(0.05\)               & \(4.2 \times 10^{-10}\) & \(6.3 \times 10^{-4}\)  \\
HD\(^+\)         & \(9.0 \times 10^{-4}\) & \(0.6\)                & \(1.2 \times 10^{-18}\) & \(7.3 \times 10^{-8}\)  \\
HeH\(^+\)        & \(1.0 \times 10^{-3}\) & \(74\)                 & \(1.7 \times 10^{-14}\) & \(1.8 \times 10^{-3}\)  \\
\(^7\)LiH        & \(2.1 \times 10^{-3}\) & \(2.36 \times 10^{4}\)  & \(9.0 \times 10^{-20}\) & \(1.7 \times 10^{-7}\)  \\
\(^7\)LiH\(^+\)  & \(3.2 \times 10^{-3}\) & \(2.34 \times 10^{3}\)  & \(4.6 \times 10^{-20}\) & \(1.4 \times 10^{-30}\) \\
H\(_3^+\)        & \(1.1 \times 10^{-3}\) & \(1.76 \times 10^{-5}\)  & \(8.0 \times 10^{-17}\)  & \(1.6 \times 10^{-12}\)  \\
H\(_2\)D\(^+\)    & \(7.7 \times 10^{-4}\) & \(0.33\)               & \(1.6 \times 10^{-20}\)  & \(1.6 \times 10^{-12}\)  \\
\hline
\end{tabular}
\end{table}

\vspace{5mm}
\footnotesize
\textbf{References:} \\
1. Wavelengths: Dalgarno et al. (1973), Oka (2004), Raich et al. (2008), Stankevich (1961), Tennyson (1984), Wolniewicz (2011), Ulivi (1991). \\
2. Einstein coefficients: Coppola et al. (2011), Gianturco et al. (1996), Goldsmith et al. (2010), Kawaoka et al. (1971), Khersonskii (1987), Moss (2002). \\
3. Relative abundances at \( z = 10 \): Galli (2013). \\


\section{Wavelength ranges of spectral distortions}
Since all non-hydrogenic spectral lines are weaker than the 21 cm hydrogen line, the latter will dominate observational efforts. We assessed whether the frequency ranges of hydrogen and other transitions overlap. Atomic hydrogen distorts the CMB at redshifts z = 6.5 to 200 (Pritchard, 2012), corresponding to present-day wavelengths of ~160 – ~4240 cm. Assuming that other species distort the CMB over the same redshift range, we identify transitions whose current wavelength ranges overlap with the calculated wavelength space.
\begin{equation*}
 \begin{cases}
   (z_{recomb}+1)\lambda_i < (z_{init}+1)\lambda_H\\
   (z_{init}+1)\lambda_i > (z_{recomb}+1)\lambda_H
 \end{cases}
\end{equation*}
The solution of this system is $\lambda\in[0.8;563]cm $. The strongest line in this range is the deuterium line.

\section{Conclusions}
The analysis presented in this work reveals that hyperfine transitions in atoms/ions and rotational transitions in molecules/ions generate signals orders of magnitude weaker than the 21 cm hydrogen hyperfine line. The dominant non-hydrogenic contribution stems from rotational transitions in the helium hydride ion (HeH$^+$), yet its amplitude remains suppressed by a factor of $\sim\!10^4$ compared to hydrogen. Within the 21 cm frequency band, the hyperfine transition of atomic deuterium (D) emerges as the most significant non-hydrogenic feature, despite being six orders of magnitude fainter than its hydrogen counterpart.

\setlength\parindent{-24pt}

\par

\end{document}